\documentclass{elsart}

\usepackage{graphicx}

\newcommand{\CMSG}{CoMnSi$_{1-x}$Ge$_{x}$}
\newcommand{\CMGS}{CoMnGe$_{1-x}$Sn$_{x}$}
\newcommand{\Mnx}{MnFeP$_{1-x}$As$_{x}$}
\newcommand{\Gdmagx}{Gd$_5$(Si$_{1-x}$Ge${_x}$)$_{4}$}
\newcommand{\Tc}{$T_{\rm{C}}$}
\newcommand{\Ts}{$T_{\rm{struct}}$}

\usepackage{amssymb}

\pdfoutput=1
\begin{document}

\begin{frontmatter}



\title{Phase diagram and magnetocaloric effect of CoMnGe$_{1-x}$Sn$_{x}$ alloys}
\author[Phys]{J.B.A. Hamer}
\author[Phys]{R. Daou}
\author[Phys]{S. \"{O}zcan}
\author[MSM]{N.D. Mathur}
\author[MSM]{D.J. Fray}
\author[MSM]{K.G. Sandeman\corauthref{cor1}}
\corauth[cor1]{Corresponding author.}
\ead{kgs20@cam.ac.uk}
\address[Phys]{Cavendish Laboratory, University of Cambridge, JJ Thompson 
Avenue, Cambridge, CB3 0HE, UK}
\address[MSM]{Dept. of Materials Science and Metallurgy, University of 
Cambridge, New Museums Site, Pembroke Street, Cambridge, CB2 3QZ, UK}

\begin{abstract}
We propose the phase diagram of a new pseudo-ternary compound, {\CMGS}, in the range ${x{\leq}0.1}$.  Our phase diagram is a result of magnetic and calometric measurements.  We demonstrate the appearance of a hysteretic magnetostructural phase transition in the range $x=0.04$ to $x=0.055$, similar to that observed in CoMnGe under hydrostatic pressure.  From magnetisation measurements, we show that the isothermal entropy change associated with the magnetostructural transition can be as high as  4.5~JK$^{-1}$kg$^{-1}$ in a field of 1~Tesla.  However, the large thermal hysteresis in this transition (${\sim}$20~K) will limit its straightforward use in a magnetocaloric device.
\end{abstract}

\end{frontmatter}

\section{Introduction}
\label{Intro}
The current resurgence of interest in magnetic refrigeration at room temperature is founded on the discovery of magnetic refrigerants with large magnetocaloric effects~\cite{pecharsky_1997a,tegus_2002a}. In addition, there are two potential environmental benefits relative to conventional gas compression refrigeration: (i) the negligible greenhouse impact of a solid magnetic refrigerant and (ii) an increase of up to 40\% in the energy efficiency of the magnetic refrigeration cycle~\cite{zimm_1998a}.  With regard to the latter, it is clear that the efficiency of the magnetic refrigeration cycle depends on the size of the magnetocaloric effect of the refrigerant.  However, it is worth noting that high efficiency also relies on the reversibility of the process of magnetising and demagnetising the refrigerant.  Such reversibility is far from guaranteed in the vicinity of a first order phase transition, where the largest magnetocaloric effects are to be found~\cite{tishin_2003a}.  

In recent times, much effort has been devoted to magnetocaloric effects associated with materials with first order, rather than continuous magnetic phase transitions in the near-room temperature range.  In inexpensive d-metal-based alloys, a first order transition is often preferred in order to counteract the diminution of the magnetocaloric effect arising from the magnetic dilution (alloying) usually required to provide a magnetic phase transition at room temperature.  Such dilution of the magnetic species can lead to a reduction of the magnetocaloric effect compared to that seen in the benchmark material, gadolinium, unless the magnetic transition in the d-metal alloy is first order~\cite{pecharsky_2001a}.  However, the presence of magnetic and thermal hysteresis, generic in practice to the dynamics of a first order phase transition, provides a potential obstacle to the use of first order magnetocaloric effects.  Hysteresis limits the efficiency of the cycle by reducing the effective cooling power of the refrigerant.

First order magnetic phase transitions examined for the purposes of magnetocaloric materials research fall broadly into two categories: magnetoelastic and magnetostructural transitions.  In the former case, a lattice distortion, stretch or compression takes place, often principally along one axis.  Examples include the Curie transition in \Mnx~\cite{tegus_2002a} and the metamagnetic transition in \CMSG~\cite{sandeman_2006a}.  In such cases, hysteresis can be minimal.  In the case of magnetostructural transitions, a different crystal structure is found on either side of the phase transition, for example in \Gdmagx~\cite{pecharsky_1997a} or in shape memory alloys such as Ni$_{2+x}$Mn$_{1-x}$Ga~\cite{pareti_2003a}.  In these cases, hysteresis can be much larger and there have been efforts, via controlled doping, to minimise it~\cite{provenzano_2004a}.

In this article we present the phase diagram of a new alloy system, \CMGS, designed to exhibit a hysteretic, magnetostructural phase transition similar to that observed in Ni$_{2+x}$Mn$_{1-x}$Ga.  We analyse the magnetocaloric effect of the material system, in particular in the region where a continuous Curie temperature and a first order martensitic structural phase transition are tuned together by chemical pressure to produce a combined, first order magnetostructural phase transition.  Our results have implications for the study of hysteretic magnetic transitions in the context of magnetocaloric materials research.

\section{Previous work on CoMnGe and CoMnSn}
\label{CMGWork}
The magnetic and structural properties of CoMnGe have been investigated by several groups~\cite{niziol_1989a,kaprzyk_1990a,johnson_1975a,kanomata_1995a,jeitschko_1975a}.  CoMnGe has a low temperature orthorhombic structure, space group {\em Pnma}, and transforms via a diffusionless, martensitic transition to a higher temperature hexagonal form, space group {\em P6$_{3}$mmc}.  The temperature, {\Ts} of this transition is highly dependent on the level of occupation of the Mn and Co sites~\cite{johnson_1975a,kanomata_1995a,koyama_2004a} .  As a result, for CoMnGe, published values of {\Ts} on heating vary from 420~K~\cite{niziol_1981a} to 650~K~\cite{kanomata_1995a}.  The magnetism of CoMnGe is based on localised Mn and Co moments, with $\mu_{Mn}\sim2.9\mu_B$ and $\mu_{Co}\sim0.9\mu_B$ in the ferromagnetic, orthorhombic groundstate~\cite{niziol_1982a}.  This state has a {\Tc} of around 345~K~\cite{niziol_1989a}. A metastable hexagonal state can be formed by quenching the material from 1200~K, and this form of the material has a lower saturation moment and a lower {\Tc} of around 283~K~\cite{kaprzyk_1990a}.  

\begin{figure}
\includegraphics[width=0.9\columnwidth]{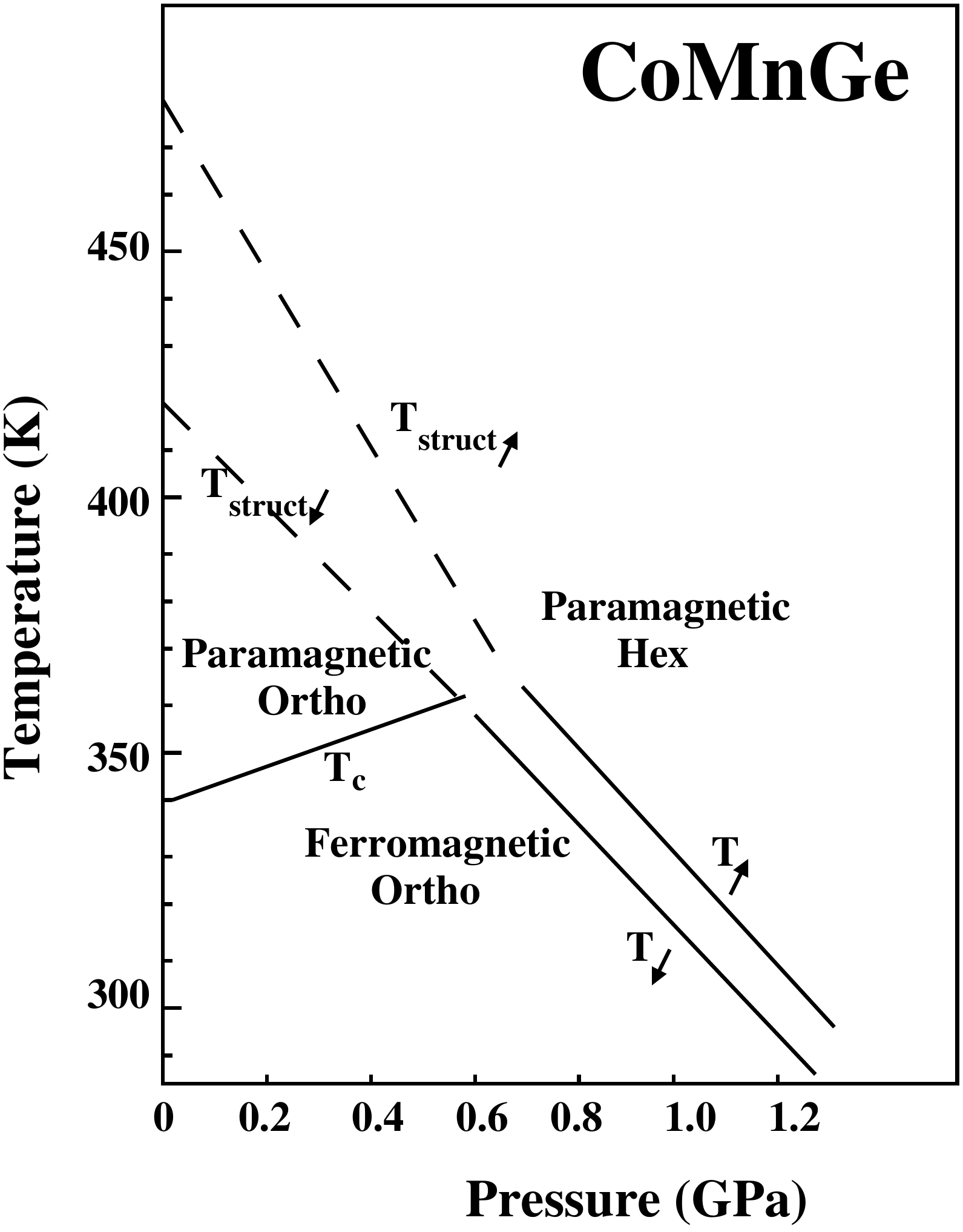}
\caption{Magnetic and structural phase diagram of CoMnGe under hydrostatic pressure, after Nizio{\l}~\cite{niziol_1983a}.  The structural and magnetic phase transitions merge at around 0.6~GPa to yield a combined, magnetostructural transition.}
\label{NiziolFig}
\end{figure}

There has been significant work on the effect of pressure on the magnetic phase diagram of orthorhombic CoMnGe.  Nizio{\l} et al. observed a simultaneous decrease of {\Ts} and increase of {\Tc} with hydrostatic pressure until a triple point was observed at a pressure of 0.6~GPa, where the two transitions merge to give a combined transition which is magnetostructural (see Fig.~\ref{NiziolFig}).  Such behaviour seems generic to this material system under pressure, as it  was shown by the same authors across the Co$_{1-x}$Ni$_{x}$MnGe series~\cite{niziol_1983a,zach_1984a} and previously in NiMnGe~\cite{anzai_1978a}.  There has been an indication of a triple point induced by Mn vacancies in samples of composition CoMn$_{1-x}$Ge~\cite{johnson_1975a} but no similar observation has been been made to our knowledge using chemical pressure via substitution. 

CoMnSn is a less studied material.  Although first reported as being a cubic ferromagnet with a {\Tc} of 670~K~\cite{castelliz_1953a}, later work has established that its groundstate has the same {\em P6$_{3}$mmc} hexagonal structure as the high temperature form of CoMnGe, and that it is a ferrimagnet with {\Tc}$\sim$145~K~\cite{bazela_1981a}.  Although no transition to a low temperature orthorhombic phase is observed, we can say nominally that  {\Ts}$<${\Tc}.

In summary,  {\Ts}$>${\Tc} in CoMnGe whereas {\Ts}$<${\Tc} in CoMnSn.  There is therefore the possibility that {\CMGS} could exhibit a combined, magnetostructural transition if {\Ts} and {\Tc} can be tuned together by chemical pressure in an analagous way to the effect of hydrostatic pressure in CoMnGe-based materials.  This is our motivation for synthesising {\CMGS}.

\section{Experimental}
\subsection{Sample preparation}
Samples of {\CMSG} with $x\leq0.1$ were prepared by induction melting pieces of elemental Mn
(99.99\%, chemically etched according to the method used by Fenstad~\cite{fenstad_2000a}), Co (99.95\%,  electropolished to the correct mass), Ge (99.9999\%) and Sn (99.9985\%) in 1 bar of argon.  Weight losses were less than 1\%.  All samples were annealed in evacuated
silica ampoules at 1023~K for 60~hours, cooled at 5~K per minute to 643~K, and finally slowly cooled at 0.1~K per minute to room temperature.  The last step ensured that the samples were slowly cooled through the structural phase transition in order to maximise the proportion of the desired orthorhombic phase.  

\subsection{Characterisation}
Room temperature X-ray diffraction of powdered samples and Rietveld refinement of lattice parameters and atomic coordinates in the orthorhombic and hexagonal phases (where present) were performed.  Measurements of magnetisation were made using a Princeton Measurements Corporation vibrating sample magnetometer (maximum field 1.8~T).  Isothermal magnetisation curves were taken with increasing magnetic field after zero field cooling from a temperature above {\Tc}.  To quantify the magnetocaloric effect in our {\CMGS} series indirectly, we use a Maxwell relation to obtain the isothermal change in total entropy from the isothermal $M(H)$ curves:
\begin{equation}
\Delta S_{\rm total}(T,\Delta H) = \int_{0}^{\Delta H}{\left(\partial 
M \over \partial T\right)_{H}}\,\,dH \, , 
\label{Maxwell}
\end{equation}
where $\Delta H$ is the maximum magnetic field applied.  Finally, to identify {\Tc} and {\Ts}, zero field heat capacity measurements were performed using a TA Instruments Q1000 differential scanning calorimeter in the temperature range 120~K to 673~K.


\section{Results and discussion}
\subsection{Structure and calorimetry}
\begin{figure}
\begin{center}
\includegraphics[width=0.9\columnwidth]{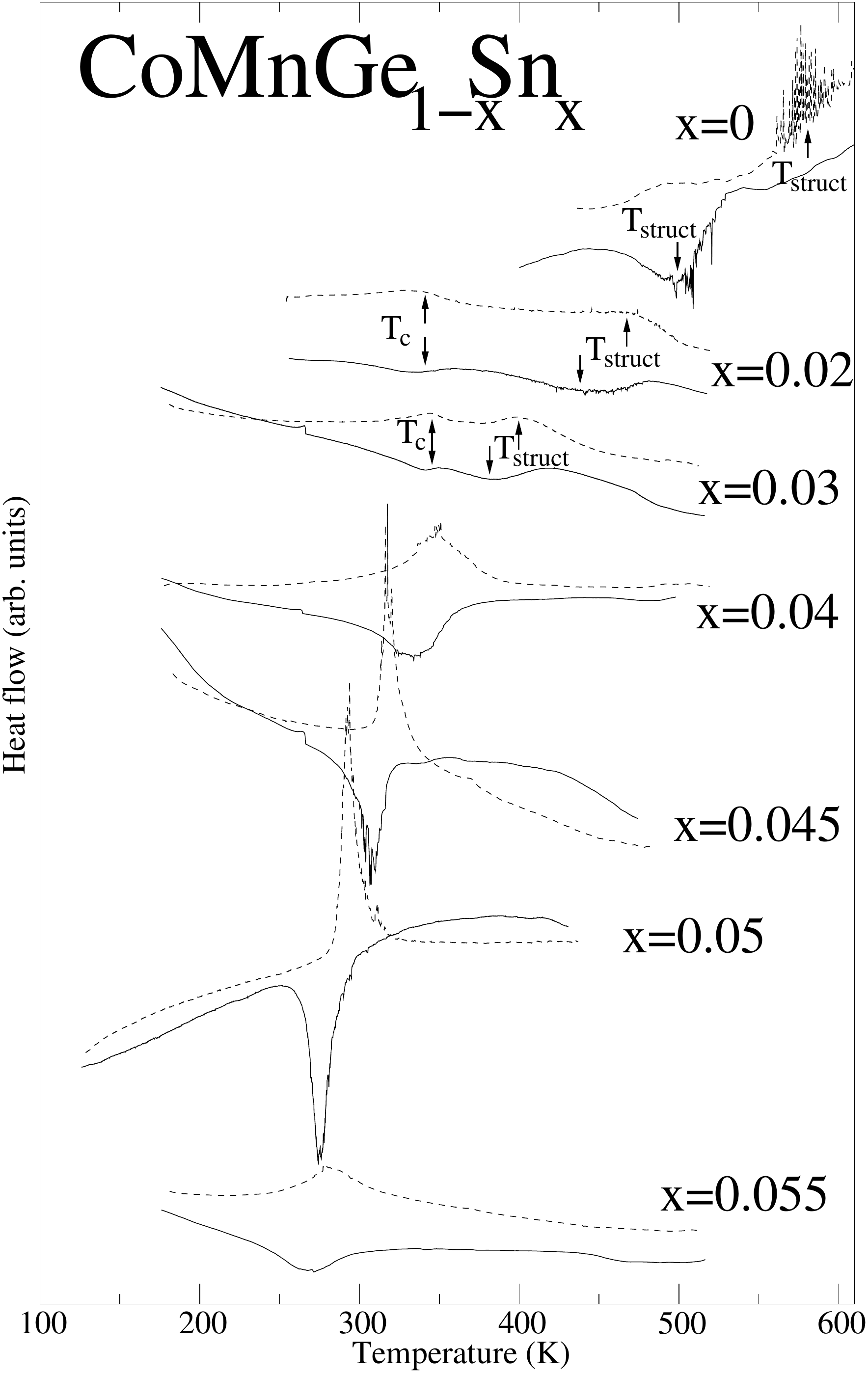}
\caption{Calorimetry data for samples of {\CMGS} with $x=0$ to $x=0.055$.  Heat flow data has been normalised per unit sample mass, but translated vertically for clarity.  Measurements of each sample were taken on heating and cooling, with the heating curve being the upper curve of each pair (dotted line).  We see that as the level of Sn substitution increases, the Curie and structural transitions merge to form a combined, magnetostructural transition.}
\label{DSCFig}
\end{center}
\end{figure}

Our X-ray diffraction data shows that the samples with $x{\leq}0.03$ are largely composed of a single orthorhombic phase, space group {\em Pnma} at room temperature. As the level of Sn substitution is increased above $x=0.03$, phase coexistence of orthorhombic and hexagonal phases occurs, until for $x>0.06$, only the hexagonal phase is present.  For samples with $x{\leq}0.03$, our calorimetry data show the presence of two peaks, where the lower peak is smaller and can be associated with the Curie temperature of the orthorhombic phase (see next section).  The higher peak is more pronounced and hysteretic, and as can be seen in Figure \ref{DSCFig}, it moves downwards in temperature and its associated hysteresis decreases with increasing Sn substitution.  We identify this highly hysteretic peak with the structural phase transition at {\Ts} between the lower temperature orthorhombic and higher temperature hexagonal structures.  Its value in CoMnGe is 580~K on heating, well within the range of values found in the literature. As $x$ increases from $x=0.04$ to $x=0.055$, the structural peak reaches the temperature of the lower (Curie) peak and the two peaks merge.  From this data, it seems as though this signals the onset of a single magnetostructural transition.  Magnetic data in the next section would seem to support this claim.  For $x{\leq}0.06$, we do not observe the {\Ts} peak in our calorimetry data, indicating that the system is hexagonal down to the lowest temperatures measured (123~K).

\subsection{Magnetic measurements}
\begin{figure}
\includegraphics[width=\columnwidth]{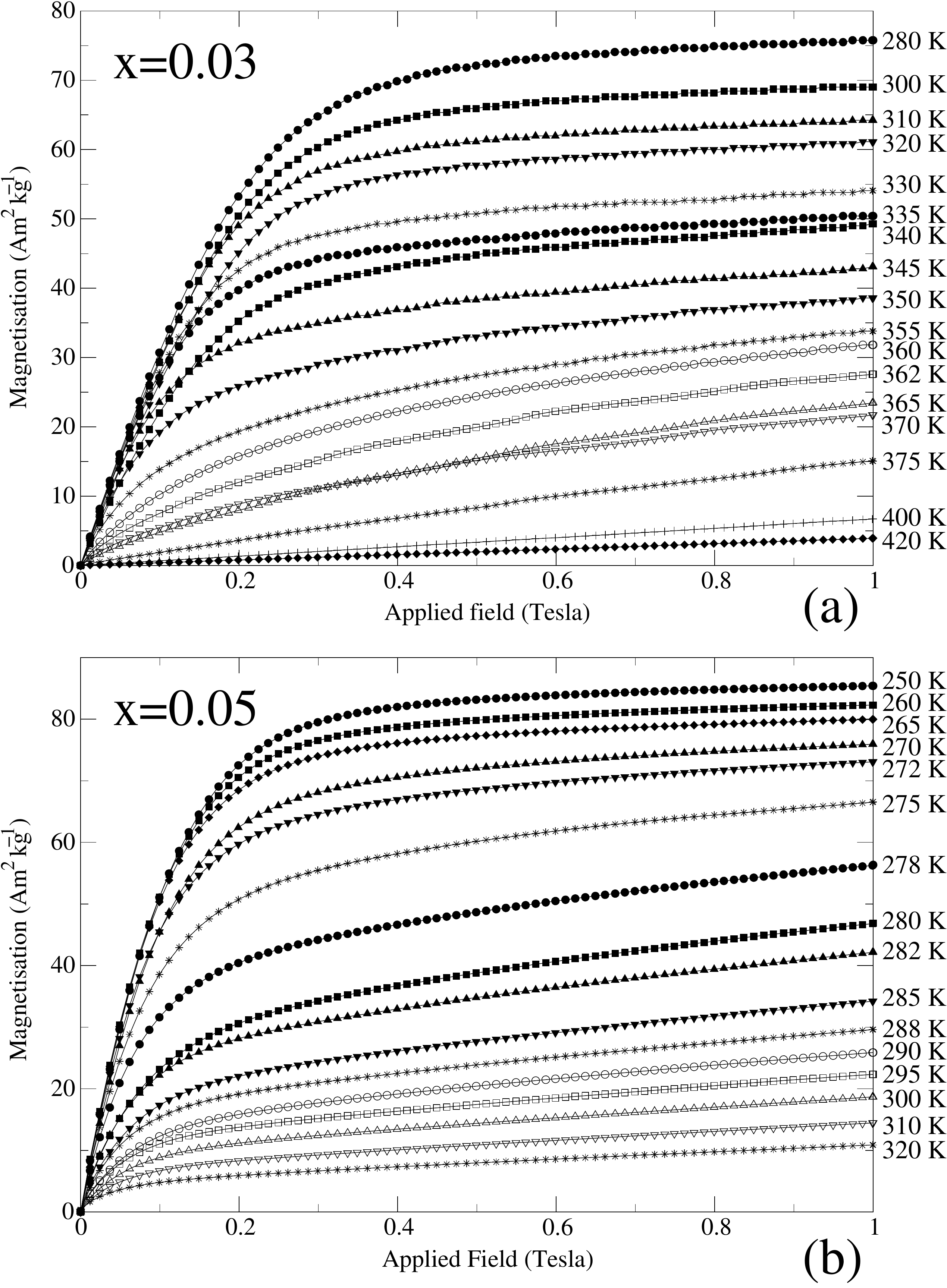}
\caption{Isothermal magnetisation curves for samples with Sn substitutions levels (a) $x=0.03$ and (b) $x=0.05$.  Data was taken after zero field cooling from above {\Tc} in each case. The $x=0.03$ sample displays the behaviour of a ferromagnet with a continuous transition at around {\Tc}=340~K.  The $x=0.05$ sample shows a more dramatic change in magnetisation with temperature, which is also reflected in calorimetry data.}
\label{MHFig}
\end{figure}

\begin{figure}
\includegraphics[width=\columnwidth]{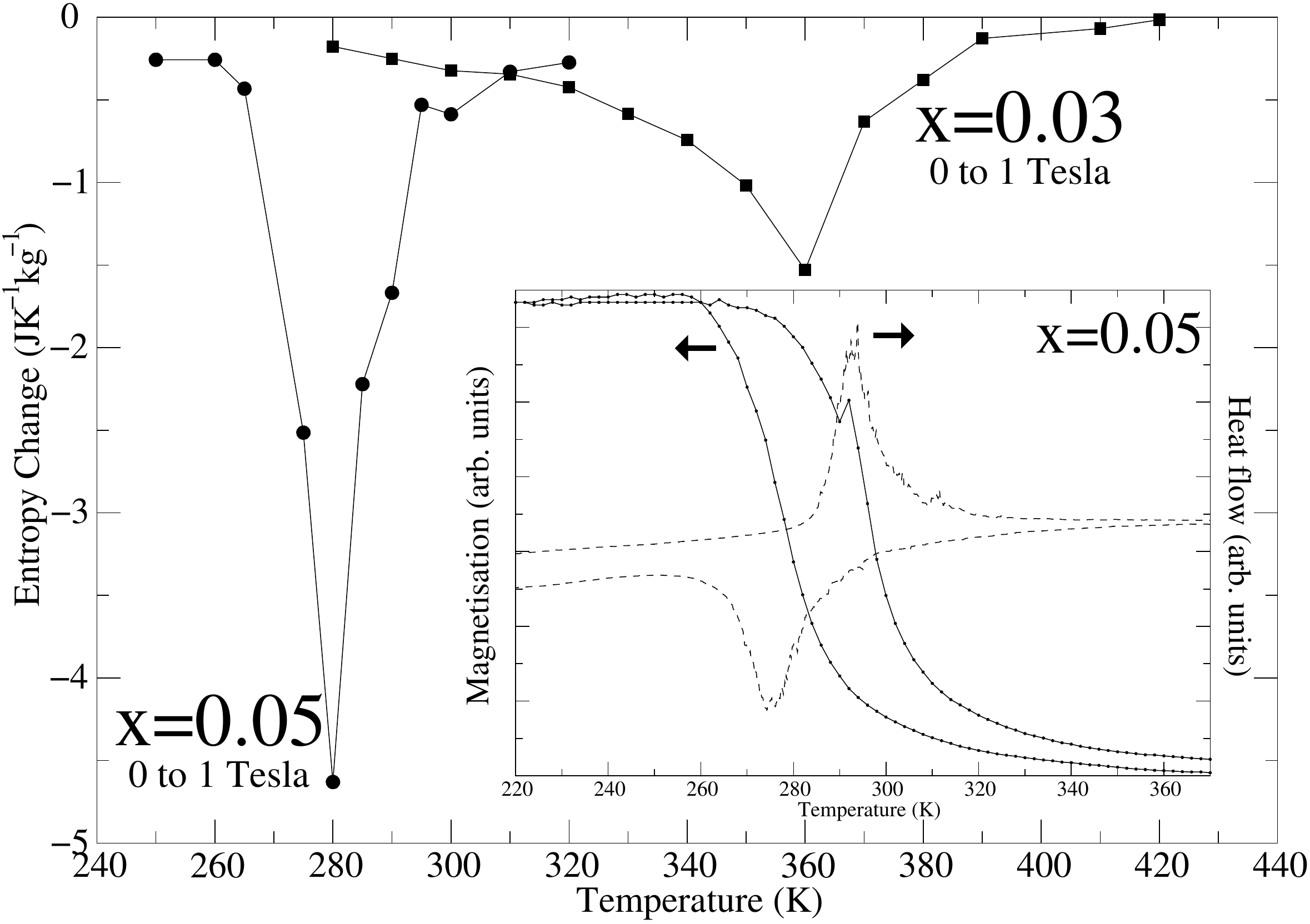}
\caption{Isothermal entropy changes for {\CMGS} with $x=0.03$ and $x=0.05$ in an applied field change of 1 Tesla.  Data is calculated using the magnetisation curves in Figure \ref{MHFig} and Eq. \ref{Maxwell}.  We see that the combined magnetostructural transition in the $x=0.05$ sample shows a significantly higher entropy change, albeit over a relatively narrow temperature window.  The inset indicates that the structural transition and the magnetic transitions are simultaneous in the $x=0.05$ sample by overlaying field-cooled magnetisation data taken in a field of 0.01~T with zero field calorimetry data.}
\label{EntropyFig}
\end{figure}

Isothermal magnetisation curves on increasing the applied magnetic field are shown in Figure \ref{MHFig} for samples with Sn substitution levels $x=0.03$ and $x=0.05$.  These samples are chosen here for their contrasting behaviour, identifiable with different regions of the {\CMGS} phase diagram which we will propose later.  The $x=0.03$ sample is representative of samples with $x{\leq}0.03$ and has a smoothly evolving magnetisation, from paramagnetism to ferromagnetism, indicative of a continuous phase transition.  In this sample, calorimetry shows that the magnetic transition is separated from the structural one (Figure~\ref{DSCFig}).  However, in the $x=0.05$ sample, which is representative of the range $x=0.04$ to $x=0.055$, the magnetisation changes much more abruptly at around 280~K.  As shown in Figure~\ref{DSCFig}, calorimetery reveals the presence of only one phase transition.  Furthermore, measurements of magnetisation with temperature in a low applied field (0.01~T) show a first order, hysteretic, phase transition, coincident with the calorimetry peak.  This is shown in the inset to Figure~\ref{EntropyFig}.  We therefore conclude that a single, combined, magnetostructural transition exists in the range $x=0.04$ to $x=0.055$.  Our samples with $x=0.06$ and $x=0.10$ showed a broad Curie transition at around 260~K, which is consistent with the continuous transitions seen in hexagonal CoMnGe-based structures~\cite{kaprzyk_1990a}.

From our isothermal magnetisation measurements, we used Eq. \ref{Maxwell} to estimate the isothermal entropy change on application of a 1~Tesla magnetic field.  The results are shown for $x=0.03$ and $0.05$ in Figure~\ref{EntropyFig}.  We note that there is only one curve for each material, as data was only taken on increasing the magnetic field.  As can be seen, the combined magnetostructural transition of the $x=0.05$ sample exhibits a peak entropy change of greater than 4.5~JK$^{-1}$kg$^{-1}$, albeit over a limited temperature window.  This peak value of entropy change may appear large in the context of other magnetocalorics in similar fields, e.g. Gd.  However two concerns with such a comparison must be made.  Firstly, {\CMGS} posseses a relatively high heat capacity per kg due to the magnetic dilution by Ge/Sn (and to a lesser extent Co). Since the adiabatic temperature change, $\Delta T$ of any material is inversely proportional to its heat capacity, such a factor serves to diminish the available $\Delta T$~\cite{pecharsky_2001a}.  Indeed this is a problem generic to d-metal-based alloys, mentioned in Section~\ref{Intro}.  

Secondly, the hysteresis which has entered the magnetic transition as a result of being combined with a first order, martensitic structural transition will have an impact on the available refrigerant capacity of the material.  
The magnetisation curves plotted in Figure~\ref{MHFig} and the resulting entropy curves in Figure~\ref{EntropyFig} are only one of two possible sets; there is also another possible set for magnetisation measured in decreasing applied field. The available refrigerant capacity, which might seem to be the integral of the isothermal entropy curve presented in Figure~\ref{EntropyFig}, is in fact decreased by $\mu_{0}\Delta H \Delta M$ in an idealised, sharp first order transition with field hysteresis $\Delta H$ and a magnetisation jump of $\Delta M$~\cite{hysteresis}.  In the case of the magnetostructual transitions presented here, thermal hysteresis is around 20~K, and we would expect field hysteresis to be similarly significant.  We can estimate the field hysteresis from the rate of shift of the transition temperature with applied field (of the order of 1~K/Tesla, similar to that in Ni-Mn-Ga shape memory martensitic alloys~\cite{pareti_2003a}).  Then a thermal hysteresis of 20~K corresponds to a field hysteresis of 20~Tesla. The implication is therefore that the useful refrigerant capacity of such a magnetostructrual phase transition is much lower than it might appear from a single entropy curve such as that presented in Figure~\ref{EntropyFig}.  For a magnetic refrigerant, it is therefore more attractive to keep the structural and magnetic transitions close in temperature, but not overlapping, as was attained by Koyama et al. and Lin et al.~\cite{koyama_2004a,lin_2006a}.  Proximity of the first order structural transition can tend to sharpen the continuous magnetic transition, and improve the entropic response, without invoking large magnetic or thermal hysteresis.

\begin{figure}
\includegraphics[width=\columnwidth]{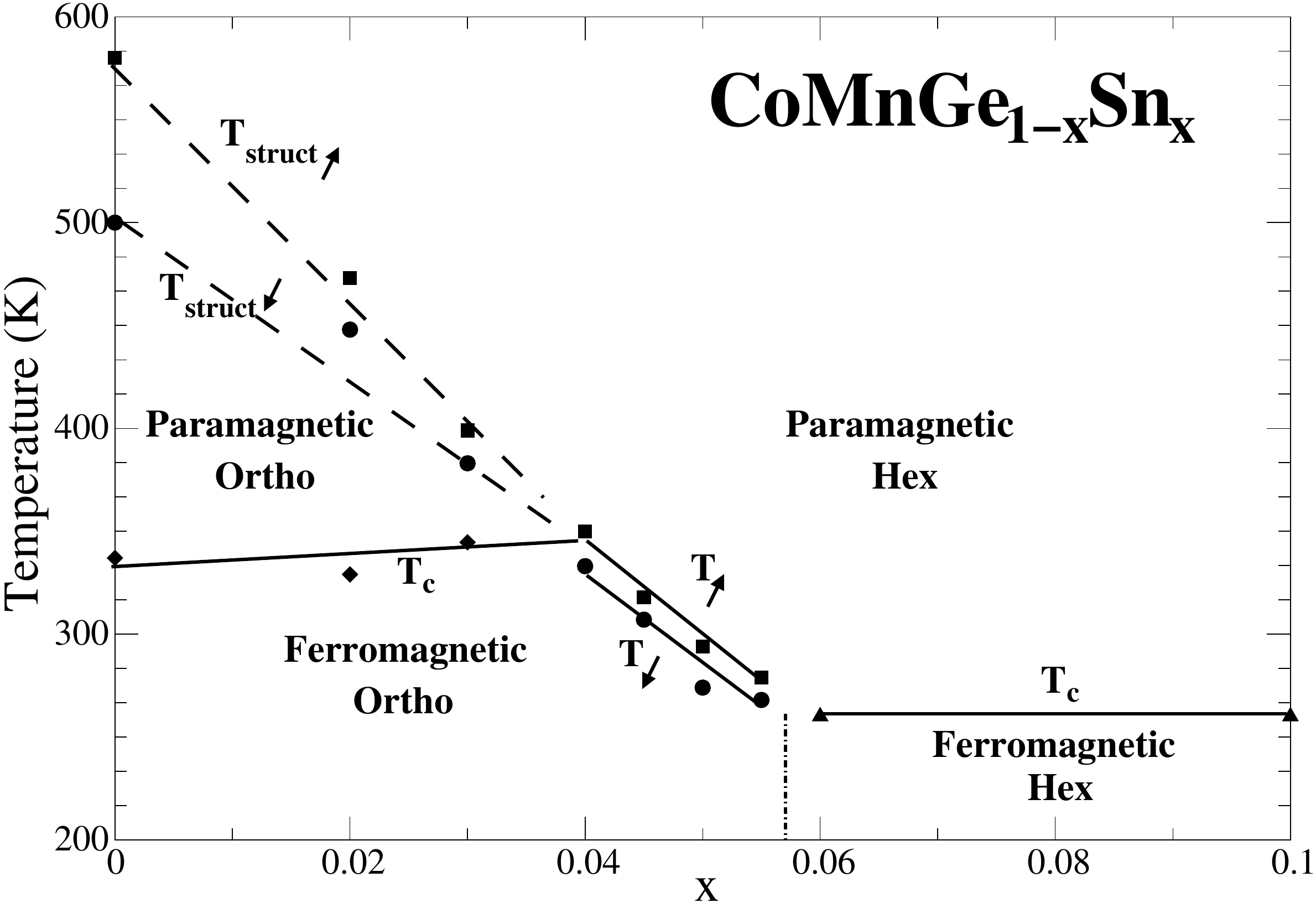}
\caption{Proposed magnetic and structural phase diagram of {\CMGS}. Diamonds indicate {\Tc} obtained from DSC measurements, while squares and circles denote {\Ts} on heating and cooling (also from DSC measurements).  For $x{\geq}0.6$, {\Tc} was obtained from magnetometry and is shown by triangles.  The phase diagram is very similar to that of CoMnGe under pressure, with the observation of a combined magnetostructural phase transition in the range $0.04{\leq}x{\leq}0.055$.  Lines are guides to the eye.}
\label{PhaseDiagFig}
\end{figure}
\section{Conclusions}
As a result of our magnetic and structural measurements, we present a phase diagram of the {\CMGS} system in Figure~\ref{PhaseDiagFig}.  It is very similar in form to that of CoMnGe under pressure (Figure \ref{NiziolFig}) and is the first observation of a triple point in a substituted CoMnGe-based material.  The orthorhombic ferromagnetic groundstate of CoMnGe gives way to the hexagonal ferrimagnetic groundstate of CoMnSn via a combined magnetostructural transition in the range $x=0.04$ to $x=0.055$.   In Figure~\ref{PhaseDiagFig}, the vertical dotted line between $x=0.055$ and $x=0.06$ indicates a nominal value of $x$ at which the orthorhombic state is no longer observed.  We also note that we can only speculate on whether the orthorhombic structural groundstate is actually ever recovered at low temperatures in the range $x>0.055$. 
From our magnetocaloric measurements, the peak isothermal entropy change in a relatively small field change of 1~Tesla is greater than 4.5~JK$^{-1}$kg$^{-1}$ for $x=0.05$.  However, the large hysteresis in the transition, brought about as a result of the martensitic structural change, probably negates any improvement in peak entropy relative to that in the continuous Curie transition ($x{\leq}0.03$) from an applied point of view.  This observation serves to illustrate the problems when considering materials with hysteretic structural phase transitions as magnetic refrigerants.

\section*{Acknowledgements}
We thank K. Roberts for help with sample preparation and G.G. Lonzarich for use of sample synthesis facilities.  NDM and KGS acknowledge financial support from The Royal Society.  The work was funded through UK EPSRC Grant GR/R72235/01.

\end{document}